\documentclass[12pt,preprint]{aastex}

\shorttitle{Kelvin$-$Helmholtz instability}
\shortauthors{Li et al.}

\begin{document}

\title{Flow instabilities in solar jets in their upstream and downstream regimes}

\author{Xiaohong Li, Jun Zhang, Shuhong Yang \& Yijun Hou}

\affil{CAS Key Laboratory of Solar Activity, National
Astronomical Observatories, Chinese Academy of Sciences, Beijing
100101, China; \email{lixiaohong@nao.cas.cn; zjun@nao.cas.cn}}

\affil{School of Astronomy and Space Science, University
of Chinese Academy of Sciences, Beijing 100049, China}

\begin{abstract}
Using the Atmospheric Imaging Assembly 304 {\AA} images obtained from
the \emph{Solar Dynamics Observatory}, we study two jets which occurred
during the M5.8 class flare on 2017 April 3 and the M5.5 class flare on
2016 July 23, respectively. During the M5.8 class flare, many vortex-like
structures occurred in the upstream and downstream regimes of the associated jet.
While the jet was ejected upwards to the corona, some dark material at its
base flowed through a bright structure with a velocity of 110 km s$^{-1}$.
The boundary between the material and the structure changed from smooth to uneven.
Later, the jet material at the higher atmosphere started to fall down
with velocities of over 200 km s$^{-1}$, and the left boundary of the jet
developed into a sawtooth pattern. The vortex-like structures
were formed, and the growth rates of two structures were presented.
During the M5.5 class flare, we also observed
many vortex-like structures in the downstream regime of the jet.
At the late stage of the jet, some material at the south boundary of the
jet fell back to the solar surface, and vortex-like structures at the boundary
grew from ripple-like minim into vortices with diameters of 3.4 $-$ 5.4 Mm.
The growth rates of the vortex-like structures were calculated.
We suggest that the vortex-like structures in the upstream
regime are the manifestations of Kelvin$-$Helmholtz instability, and those in
the downstream regime are simultaneously driven by Kelvin$-$Helmholtz
instability and Raleigh$-$Taylor instability.

\end{abstract}


\section{Introduction}

Raleigh$-$Taylor instability (RTI) and Kelvin$-$Helmholtz
instability (KHI) are basic instabilities in fluids and magnetized plasma.
The RTI occurs at the interface between two fluids of different
densities whenever fluids experience a pressure gradient that
opposes the density gradient \citep{Tay1950, Sha1984}. The RTI
is mostly gravitationally driven such as a dense fluid is supported
against gravity above a lighter fluid. The influence of a
magnetic field on the RTI depends on its component parallels
to the interface, which can suppress the growth of the magnetic
RTI through magnetic tension. Observations of the RTI include
mushroom clouds from atmospheric nuclear explosions, supernova
explosions in which expanding core
gas is accelerated into denser interstellar medium \citep{Wang2001},
and the finger-like structures in Crab Nebula \citep{Hes1996}.

The KHI arises at the interface of two parallel flows. In non-viscous
fluids, the KHI will occur as long as there is a velocity shear.
Since viscosity and magnetic field have stabilizing influences, a
threshold of the velocity difference is required for the KHI to
take place in magnetized plasma \citep{Chan1961}. In astrophysics
and space physics, the KHI has been observed in many active phenomena,
e.g., the solar wind \citep{Suess2009}, Earth's magnetopause \citep{Hase2004},
planetary magnetotails \citep{Mas2010}, and cometary tails \citep{Ersh1980}.

Solar atmosphere is made of hot and almost fully ionized plasma.
The differences in densities and flow speeds between an
expanse of erupting plasma and the background plasma may
trigger the RTI and KHI. The RTI in the solar corona
has been studied through the formation of plumes when a relatively
dense solar prominence overlies a less dense plasma bubble \citep{Ryu2010, Ber2011}.
The fragmentation of prominence eruption as the material falls
back to the solar surface \citep{Inn2012, Car2014} and the filamentary
structure associated with emerging magnetic flux \citep{Iso2005}
are manifestations of the RTI as well. These observations are
in accordance with the numerical simulations that in three dimensions
the RTI results in the formation of finger-like structures
elongated in the direction of the magnetic field \citep{Sto2007}.
The KHI is also believed to operate in the solar atmosphere, with velocity
difference exceeds twice the order of Alfv$\acute{e}$n velocity.
The KHI can be identified by the appearance of growing ripples or the
vortices which form across the boundary between two flows.
\cite{Fou2011} observed the KHI in a fast coronal mass ejection
(CME) event using the data taken by the \emph{Solar Dynamics
Observatory} \citep[\emph{SDO};][]{Pes2012}. \cite{Ofm2011}
confirmed the occurrence of the KHI in the solar corona, and
the KHI in the solar prominence was also investigated by
\cite{Ber2010} and \cite{Ryu2010}. \cite{Li2018} reported that
the KHI developed at the boundary of a jet due to the strong
velocity shear ($\sim$ 204 km s$^{-1}$) between two flows.

Solar jets, the plasma ejections along the open magnetic field
lines in the solar corona, were discovered in the 1980s and then
studied after the launch of the Japanese \emph{Yohkon} satellite
in the 1990s \citep{Shib1994, Shim1996}. Subsequently, the unprecedented
high-resolution observations by the \emph{Hinode} \citep{Kos2007},
\emph{Solar TErrestrial RElations Observatory} \citep[\emph{STEREO};][]{
Kai2008}, \emph{SDO}, and \emph{Interface Region Imaging Spectrograph}
\citep[\emph{IRIS};][]{DeP2014} have made great improvements in our understanding
of the solar jets \citep{Sav2007, Shen2011, Tian2014}. Solar jets
have been observed all over the solar atmosphere including coronal holes \citep{Yang2011},
quiet regions \citep{Hong2011} and active regions \citep{Li2015}. Many wavelengths, e.g.,
X-ray \citep{Ster2015}, extreme ultraviolet \citep[EUV;][]{Chae1999} and
H$\alpha$ \citep{Yoko1995} are employed to detect the jets.
Coronal jets have apparent lengths of 10 $-$ 400 Mm and widths of
5 $-$ 100 Mm. The speeds of jets range from 10 to 1000 km s$^{-1}$, with a
mean value of 200 km s$^{-1}$ \citep{Shim1996}. The lifetimes of jets are
10 $-$ 70 min, with a median value of 20 $-$ 40 min \citep{Nis2009}.
Jets have temperature in the range of 0.05 to 2.0 MK, while
electron densities have been reported as 6.6 $\times$ 10$^{9}$ to
3.4 $\times$ 10$^{10}$ cm$^{-3}$ \citep{Yan2011}. The magnetic field
strength of the jets is a few Gauss \citep{Puc2013}. The morphology,
formation mechanisms and dynamic characters of the jets, as well
as their relations to other coronal structures, have been carefully
studied \citep{Rao2016}. However, studies of the instabilities in
solar jets are still in early stages.

In this study, using the high-resolution data from the \emph{SDO},
we present observations of the vortex-like structures in two solar jets.
The first jet was located at active region (AR) 12644 and was associated
with an M5.8 class flare on 2017 April 3. During this jet, we
observed the vortex-like structures both in its upstream and downstream
regimes, i.e., when the jet was ejected upwards to the corona and fell down
from the higher atmosphere. We also observed the vortex-like structures
driven by the falling material of the jet during the M5.5 class flare on
2016 July 23. The vortex-like structures in the upstream regime of the jet
are interpreted as evidence of the KHI, and the vortex-like structures in
the downstream regime may be caused by both the RTI and KHI.

\section{Observations and data analysis}

We adopted the Atmospheric Imaging Assembly \citep[AIA;][]{Lemen2012}
multi-wavelength images and the Helioseismic and Magnetic
Imager \citep[HMI;][]{Scher2012, Schou2012} data on board
the \emph{SDO}. For the M5.8 class flare on 2017 April 3, we chose the
AIA 304 {\AA} images, obtained from 14:00 UT to 16:00 UT with a pixel
size of 0$\arcsec$.6 and a cadence of 12 s. We also used the intensitygram
from the HMI, with a spatial sampling of 0$\arcsec$.5 pixel$^{-1}$ and
a 15 min cadence, i.e., one frame in twenty, from 2017 April 3 00:00 UT
to 2017 April 4 00:00 UT. For the M5.5 class flare on 2016 July 23, we
employed the AIA 304 {\AA} images from 05:30 UT to 06:30 UT with a pixel size of
0$\arcsec$.6 and a cadence of 12 s. In addition, we also employed the
\emph{Geostationary Operational Environmental Satellite (GOES)} data with
the 1 min cadence to examine the variation of soft X-ray 1$-$8 {\AA}
flux.

\section{Results}

On 3 April 2017, there were five ARs on the solar disk, and
AR 12644 we focus on was located at the west boundary. At 14:19 UT,
an M5.8 class flare took place in this AR. Associated with the M5.8 class
flare, a jet occurred on the north side and its brief evolution is shown
in Figure 1. The HMI intensitygram in panel (a) shows the location and
appearance of the sunspot of AR 12644. The \emph{GOES} soft X-ray 1$-$8 {\AA}
flux (see green curve in panel (b)) shows that the M5.8 class flare
reached its peak at 14:29 UT, and the appearances of the flare and
the jet at this moment are shown in panel (b). The jet took place at
about 14:20 UT, and we observed numerous vortex-like structures during
its evolution. The green rectangle in panel (c) and square in panel (d)
display the positions where the vortex-like structures are observed,
respectively.

Figure 2 shows the development of vortex-like structures which happened
near the base of the jet. At 14:30 UT, the main mass of the jet was
ejected outwards. At the base of the jet, some dark material flowed
through the bright structure with a velocity of 110 km s$^{-1}$, and
the boundary between the material and the structure was smooth as shown
in panel (a). Then, the boundary became distorted as denoted by the blue curves in
panels (b)$-$(d). The biggest distortion in each panel is indicated by the blue
arrow, with a size scale of 1.7 $-$ 2.5 Mm. The green arrows in panel (e) denote
several vortex-like structures, and the distance between two nearby structures is
roughly 2.7 $-$ 3.8 Mm. The vortex-like structures such as knots,
growing ripples and so on are common phenomena in the solar atmosphere
\citep{Sev1953, Rot1955, Sak1976, Ofm2011}. Some researchers interpreted the
vortex-like structures appearing on the boundary between the jet and
the ambient background as evidence of the KHI \citep{Li2018, Zhel2018}.
Here, the velocity shear between the dark material and the bright structure
drove the KHI, and the vortex-like structures were formed consequently.
As the KHI developed, these structures became turbulent (see panels (f) and (g)).
After the KHI, the boundary began smooth again as shown in panel (h).

Figure 3 and Movie1 show the evolution of the vortex-like structures at the
left boundary of the jet. Due to the gravity, the jet material whose speed
didn't reach the escape speed started to drop at about 15:14 UT from the higher
atmospheric layer. Along the left boundary of the jet (see slice ``A$-$B" in
panel (f)), we made a slice and the space-time plot is displayed in panel (h).
We chose some trajectories and determined that the falling velocities of the
jet material were approximately 224 $-$ 289 km s$^{-1}$.
The left boundary of the jet was smooth initially, as shown in panel (a).
At about 15:25 UT (see panel (b)), the boundary began to display ripple structures.
These structures gradually developed and grew up into vortex-like structures
which are displayed in panels (c) and (d). We chose some significant structures
(see the green rectangle in panel (e)) when they were biggest and enlarged these
structures in panel (g). The deformations were roughly 6.4 $-$ 8.8 Mm.
We chose two structures in panel (g) and measured their deformations
over time. The deformations in three minutes are plotted in Figure 4.
The deformations were exponential growths d = d$_0$ e$^{\gamma t}$, and the
growth rates $\gamma$ were estimated to be 0.0097 and 0.0088.

As the falling jet material was denser than the surrounding corona,
and the interface was at some angle to gravity (the green arrows in Figure 3(b)
display the directions of gravity at the points where the arrows start), these
vortex-like structures in the downstream regime of the jet may be caused by the RTI.
When magnetic field is not involved, the growth rate of the RTI is give as
\citep[e.g.,][]{Chan1961, Ryu2010}
\begin{eqnarray}
\gamma = \sqrt{(\rho_u - \rho_l) g k / (\rho_u + \rho_l)}
\end{eqnarray}
where k is the wavenumber, g is the acceleration due to gravity, $\rho_u$ is
the upper density and $\rho_l$ is the lower density (relative to the direction
of gravity). We take the electron number density of the surrounding environment
to be equal to $n_l$ = 10$^{15}$ m$^{-3}$, and the electron number density of
the jet is $n_u$ = 10$^{16}$ m$^{-3}$, then
the density can be derived as $\rho$ = $n$ $\cdot$ $m$ ($m$ = 1.673 $\times$ 10$^{-27}$ kg).
The solar gravitational acceleration is 273.2 m s$^{-2}$, then g is 236.6 m s$^{-2}$
considering the angle between the gravity and the interface as 60 degrees.
The distance between two structures is regarded as the characteristic
wavelength ($\sim$ 5000 km). Thus, we can estimate that the growth rate
is 0.0156 when the RTI is pure gravitationally driven. In solar jets,
the magnetic field constrains the motion of the plasma. The inclusion
of magnetic field adds a criterion for the occurrence of the RTI \citep{Chan1961, Sha1984}:
\begin{eqnarray}
k g (\rho_u - \rho_l)/ (\rho_u + \rho_l) > [(k \cdot \vec{B_u})^2 + (k \cdot \vec{B_l})^2] / \mu_0 (\rho_u + \rho_l)
\end{eqnarray}
where $\vec{B_u}$ and $\vec{B_l}$ are the magnetic intensity of the upper and lower
regions, respectively. \cite{Chen2012} assumed that the magnetic flux across
the transverse section of the jet would remain constant and found that the
magnetic field inside the jet gradually decreases with the height from
15 $\pm$ 4 G to about 3 $\pm$ 1 G at a height of 7 $\times$ 10$^4$ km ($\sim$ 100$\arcsec$).
Here, the heights where the vortex-like structures occurred were more
than 100$\arcsec$, so we assumed the magnetic field parallel to the
interface to be $B_u$ = $B_l$ = 3 G. In consequence, the left term
in equation (2) is approximately 2.43 $\times$ 10$^{-4}$, and the
right term is 1.23 $\times$ 10$^{-2}$, under which condition the
RTI won't happen. As seen from the formula, the occurrence of the RTI
would be influenced by the density and the magnetic field intensity.
When the density of the jet $\rho_u$ increases 10 times, the first term
changes a little while the second term becomes approximately one-tenth
of what it is now. So in a magnetohydrodynamic environment like corona,
the RTI would happen when the density is high enough (n $>$ 10$^{18}$ m$^{-3}$),
that's the reason why so far very few case of filament eruption
observed the RTI. When the magnetic field intensity becomes $B_u$ = $B_l$ = 2 G,
the right term would change to four-ninths of the previous value
($\sim$ 5.5 $\times$ 10$^{-3}$).
The density contrast of solar jets relative to background is
not adequate for the RTI to develop, so other reasons may also
contribute to the formation of the vortex-like structures. In our case,
as there were velocity differences (224 $-$ 289 km s$^{-1}$) between
the dropping jet material and the ambient corona, the occurrence
of the vortex-like structures may be driven by the KHI as well.
When the gravity (the RTI) is not considered, the onset condition
of the pure KHI in magnetized incompressible ideal plasma can be
deduced by \citep{Chan1961, Cow1976}
\begin{eqnarray}
(\vec{k} \cdot \vec{V_u} - \vec{k} \cdot \vec{V_l})^2 > (\rho_u + \rho_l)[(\vec{k} \cdot \vec{B_u})^2 + (\vec{k} \cdot \vec{B_l})^2]/\mu_0 \rho_u \rho_l
\end{eqnarray}
where $\vec{k}$, $\vec{V}$, $\vec{B}$, $\rho$ are the wave vector, velocity,
magnetic field intensity and mass density in the flux tube, respectively.
The subscripts `u' (`l') indicate the quantities of the upper (lower) region.
Presuming that $\vec{k}$ $\parallel$ $\vec{V_u}$ $\parallel$ $\vec{V_l}$ $\parallel$ $\vec{B_u}$ $\parallel$ $\vec{B_l}$,
then the velocity difference threshold is
\begin{eqnarray}
\vartriangle V_s = \mid \vec{V_u} - \vec{V_l} \mid = \sqrt{(\rho_u + \rho_l)(B_u ^2 + B_l ^2)/\mu_0 \rho_u \rho_l}.
\end{eqnarray}
Substituting the values into the formula, we can estimate that the velocity
difference threshold is 307 km s$^{-1}$. When the magnetic field intensity
decreases to $B_u$ = $B_l$ = 2 G, the velocity difference threshold
decreases to 205 km s$^{-1}$ accordingly, and the KHI would happen.
When the gravity is included, the instability would happen if \citep{Chan1961}
\begin{eqnarray}
\vartriangle V > \sqrt{(\rho_u + \rho_l)(B_u ^2 + B_l ^2)/\mu_0 \rho_u \rho_l - (\rho_u + \rho_l)(\rho_u - \rho_l) g / k \rho_u \rho_l}.
\end{eqnarray}
If $\rho_u$ $<$ $\rho_l$, the second term beneath the radical sign
will largen the velocity difference threshold, which means that the
gravity with density gradient suppresses the KHI. If $\rho_u$ $>$ $\rho_l$,
the second term will diminish the velocity difference threshold,
which can be regarded as the manifestation of the RTI.
Under our previous assumptions, the velocity difference threshold
declines slightly (304 km s$^{-1}$ when $B_u$ = $B_l$ = 3 G and 201 km s$^{-1}$
when $B_u$ = $B_l$ = 2 G), and the second term below the radical
sign is roughly one-fiftieth of the first term, implying that the
influence of the gravity may be very small. As the practical situation
is complex, it is difficult to decide which instability is dominant.
Hence, we interpret the occurrence of vortex-like structures as the
result of both the RTI and the KHI. Under this circumstance, the growth
rate (the imaginary part of frequency) can be deduced by \citep{Gug2010, Ber2017}
\begin{eqnarray}
\gamma = \sqrt{k g (\rho_u - \rho_l) / (\rho_u + \rho_l) - k^2 (B_u ^2 + B_l ^2) / \mu_0 (\rho_u + \rho_l) + k^2 \rho_u \rho_l (\vartriangle V)^2 / (\rho_u + \rho_l)^2}.
\end{eqnarray}
The growth rate is sensitive to changes in magnetic field intensity.
Choosing $\vartriangle$ V = 250 km s$^{-1}$, when $B_u$ = $B_l$ = 3 G,
the instability won't take place. When $B_u$ = $B_l$ = 2 G,
the growth rate is roughly 0.054, much higher than we estimated.
Using the measured growth rate ($\sim$ 0.009), we can estimate
that the magnetic field intensity is approximately 2.5 G ($B_u$ = $B_l$),
which is consistent with the actual magnetic field intensity
of the jet and the corona \citep{Chen2012}.

We also observed the vortex-like structures caused by falling material in another jet
which occurred on July 23, 2016. This jet took place in AR 12565 which was located
on the west side of the solar disk. At 05:27 UT, an M5.5 class flare occurred,
accompanying a large scale jet (see Figure 5(a)). Massive plasma was ejected outwards.
Later, some plasma fell back to the solar surface, and vortex-like structures
developed at the south boundary of the jet as shown in panels (b1)$-$(b5) and Movie2.
The south boundary of the jet was smooth at first (see panel (b1)). At 05:48 UT,
the boundary started to have ripple structures, which are indicated by the cyan
arrows in panel (b2). The average distance between two adjacent structures was
roughly 10 Mm. We studied the plasma movement at the positions of slices
``A$-$B" and ``C$-$D" (see panel (b4)), and the temporal evolutions are displayed
in Figures 6(a) and (b), respectively. At around 05:48 UT (see the vertical dashed
lines in Figures 6(a) and (b)), the plasma at the south boundary of the jet started
to fall back from the higher solar atmospheric layer. We chose several representative
structures and measured their velocities, with values from 110 km s$^{-1}$ to more
than 270 km s$^{-1}$. Same as the last jet example, the gravity
opposed the density gradient of the jet and the ambient corona,
and was at some angle to their interface (see the green arrows in Figure 5(b2)
which display the directions of gravity at the start points). Also, there
existed velocity shear between the falling material and the ambient
corona, so the vortex-like structures began to grow at the south boundary
driven by both the RTI and the KHI. We chose three isolated structures
(indicated by arrow ``1" in Figure 5(b3) and arrows ``2" and ``3" in Figure 5(b5)),
and their evolutions are presented in Figure 7.

Figure 7(a) is the space-time stack plot along the green enclosed area in
Figure 5(b3) from 05:50 UT to 05:52 UT. The cadence of AIA 304 {\AA} data is 12 s,
so there are ten images in two minutes. The blue arrows denote a vortex-like
structure (indicated by arrow ``1" in Figure 5(b3)), and its morphological changes
and position movements are displayed. In the first image of panel (a), structure ``1"
looks like a ripple. Along with the development of the KHI over time, the structure
grew and became a vortex-like structure. The diameter of structure ``1", marked by
the cyan lines and arrows, is about 4.2 Mm in the last image of panel (a).
Using the method displayed in Figure 4, we measured the changes of
the vortex size (diameter) over time and determined the growth rate $\gamma$
to be roughly 0.0087. Along the boundary of the jet, structure ``1"
moved approximately 21.8 Mm ($\sim$ 30$\arcsec$) in two minutes, thus the
velocity of structure ``1" was about 181 km s$^{-1}$. Figure 7(b) is the space-time
stack plot along the green enclosed area in Figure 5(b3) from 06:02 UT to 06:06 UT
with a cadence of 24 s. The white arrows indicate the vortex-like structure
pointed by arrow ``2" in Figure 5(b5). In 168 s, structure ``2" developed to
a vortex-like structure with a diameter of 3.4 Mm (the cyan lines and arrows
in the seventh image) and moved 21.0 Mm ($\sim$ 29$\arcsec$),
with the velocity of 125 km s$^{-1}$. The growth rate of structure ``2" was
estimated to be about 0.0070. There is another structure indicated by
the green arrows in Figure 7(b) and arrow ``3" in Figure 5(b5). This structure
was tiny at first and grew to a clear vortex in the end, with the diameter
of 5.4 Mm (see the cyan lines and arrows in the last image).
The growth rate of structure ``3" was approximately 0.0068.
The position of structure ``3" changed 32.6 Mm ($\sim$ 45$\arcsec$) in four
minutes, and the velocity of this structure was roughly 136 km s$^{-1}$.
As demonstrated before, the growth rate can be estimated using equation (6)
when both the RTI and KHI are considered. Here, we use the same assumption
that $B_u$ = $B_l$ = 2 G, $n_u$ = 10$^{16}$ m$^{-3}$ and $n_l$ = 10$^{15}$ m$^{-3}$.
The distance between two adjacent structures ($\sim$ 10 Mm) was regarded as
wavelength. Considering the angle between the gravity and the
interface as 45 degrees and the velocity difference as
$\vartriangle V$ = 200 km s$^{-1}$, then the growth rate is estimated
to be 0.0057. Remarkably, the measured values (0.0068 $-$ 0.0087) are
of the same order of the theoretically estimated values, implying that
these vortex-like structures are simultaneously driven by the KHI and RTI.

\section{Conclusions and Discussion}

With the \emph{SDO} observations, we studied the M5.8 class flare and the
associated jet in AR 12644. There occurred many vortex-like structures
during the development of the jet. As the jet material was ejected to the corona,
dark material flowed through the bright structure at the base of the jet with
velocity of 110 km s$^{-1}$, causing the KHI. The boundary between dark material
and bright structure became distorted, and vortex-like structures grew with the
biggest distortion of 1.7 $-$ 2.5 Mm. Due to the gravity, the jet material whose speed
didn't reach the escape speed dropped down from the high atmospheric layer. The
jet was denser and over 200 km s$^{-1}$ faster than the ambient corona, thus the
RTI and KHI occurred, and the boundary which was smooth at first started to
display vortex-like structures with distortions of 6.4 $-$ 8.8 Mm.
We also studied the M5.5 class flare on 2016 July 23, and at the south boundary
of the concomitant jet, the vortex-like structures also generated because of
falling material. The south boundary was smooth initially. Due to the density
difference and velocity shear ($\sim$ 110 km s$^{-1}$ to more than 200 km s$^{-1}$)
between the jet and the background, the RTI and KHI took place, and the
boundary became distorted. We analysed the evolution of the boundary,
and there were small structures grew from ripple-like minim into vortices
whose diameters were roughly 3.4 $-$ 5.4 Mm. The growth rates of the
structures were approximately 0.0068 $-$ 0.0087, and the downward velocities
of these structures were over 100 km s$^{-1}$.

In the upstream regime of the jet, the vortex-like structures such
as blobs appearing on the boundary between the jet and the corona
have been reported before \citep{Li2018, Zhel2018, Bogd2018}, and they
are interpreted as evidence of the KHI. The theory of the KHI
in solar jets has been developed in recent years, and Magnetohydrodynamic (MHD)
simulations support the presence of the KHI in solar jet \citep{Zaqa2015, Kur2016}.
Considering the distance between two structures (2.7 $-$ 3.8 Mm)
as wavelength ($\sim$ 3000 km), substituting the value $B_u$ = $B_l$ = 3 G,
$n_u$ = 10$^{16}$ m$^{-3}$ and $n_l$ = 10$^{15}$ m$^{-3}$ into equation (5),
we can estimate that the velocity difference threshold of the KHI when
the jet was ejected upwards was roughly 309 km s$^{-1}$, a little larger than
the value when the gravity is not involved (307 km s$^{-1}$). Here, the KHI occurred
when the velocity difference was 110 km s$^{-1}$, and there may be many reasons.
First of all, the density of the bright structure may be larger than that
of the corona. If we change the density of the bright structure to be
$n_l$ = 10$^{16}$ m$^{-3}$, then the velocity threshold becomes 131 km s$^{-1}$
accordingly. From the results, we can see that density contrast has a huge impact
on the onset of the KHI. What's more, the estimation is established under many
preconditions, e.g., the plasma is incompressible, ideal and $\vec{k}$ $\parallel$ $\vec{V_u}$ $\parallel$ $\vec{V_l}$ $\parallel$ $\vec{B_u}$ $\parallel$ $\vec{B_l}$, thus many parameters
can influence the occurrence of the KHI besides the velocity and density contrast.
As seen from equation (3), the effect of a magnetic field on the KHI depends
on both its intensity and orientation. Only magnetic field component parallels
to the interface discontinuity can exert a restoring force and suppresses the
growth of the KHI, thus the magnetic field configurations of the jet and
surrounding environment may affect the occurrence of KHI.
\cite{Zhel2015} investigated the KHI in surges (cool jets) by
modelling the surge as a moving twisted magnetic flux tube in
homologous and twisted magnetic field. Their numerical studies
showed that KHI occurred in both magnetic field configurations
for MHD waves propagating in axial direction, and the critical
velocity for emerging KHI was remarkably lower (24 $-$ 60 km s$^{-1}$)
when both magnetic field was twisted. Also, the compressibility
of the plasma may change the instability criteria and growth
rates \citep{Sen1964}. Regarding EUV jets as a vertically moving
flux tube (untwisted and weakly twisted), \cite{Zhel2016} found
that the critical jet velocity was 112 km s$^{-1}$ when jet was
assumed to be compressible plasma, and when the jet and
its environment were treated as incompressible, the critical velocity
became 114.8 km s$^{-1}$. Their work also proved that a weak twist of
the magnetic field in the same approximation may decrease the threshold.
What's more, viscosity may have a destabilizing influence when the
viscosity coefficient takes different values at the two sides of the
discontinuity, and therefore decrease the criteria of KHI \citep{Rud1996}.
All these results reveal that the criteria of KHI can be reduced
by many factors, and KHI can happen under the velocity difference
of 110 km s$^{-1}$ in the upstream regime of the jet.

Here, we also report the vortex-like structures occurred
when the jet material fell down, and these structures are interpreted
as the result of the KHI and RTI. The coexistence of the RTI and KHI is not
rare. In the nonlinear evolution of the RTI, the secondary KHI can
be triggered as a result of the shear flows that develop between the falling
material and the background \citep{Cat1988}. Similarly, in the nonlinear
evolution of the KHI, the rolled-up KH vortices would generate centrifugal
force and create the conditions for the development of the RTI. In a more
general case, there exist both the RTI and the KHI at the very beginning
of a real system \citep{Ye2011, Hill2018}.
In the downstream regime of the jet, the density and velocity differences
between the jet material and the corona may trigger the RTI and KHI, as long
as the stabilizing effect of the magnetic field are surpassed. Through
theoretical analysis, the coaction of the RTI and the KHI in solar jet are proved,
and there is little difference between the observed growth rates with the
theoretical ones. The combination of the RTI and KHI has been studied
before theoretically and observationally, which is so-called
coupled KH-RT instability or combined RT-KH instability \citep{Far1998, Ber2017}.
As we demonstrated before, there exist many preconditions in the derivation,
and the occurrence of the KHI can be influenced by many factors.
These factors such as the twisting of the magnetic field, the compressibility
and viscosity of the fluids, and ion-neutral collisions would
influence the RTI as well \citep{Sto2007, Libe2009, Diaz2012, Hill2016}.

\acknowledgments {
We thank the referee for valuable suggestions.
This work is supported by the National Natural Science Foundations
of China (11533008, 11790304, 11673035, 11773039, 11673034, 11790300),
Key Programs of the Chinese Academy of Sciences (QYZDJ-SSW-SLH050),
and the Youth Innovation Promotion Association of CAS (2014043).
The data are used courtesy of HMI, AIA and \emph{GOES} science teams. \\}

{}

\begin{figure*}
\centering
\includegraphics
[bb=48 170 525 625,clip,angle=0,width=1.0\textwidth]{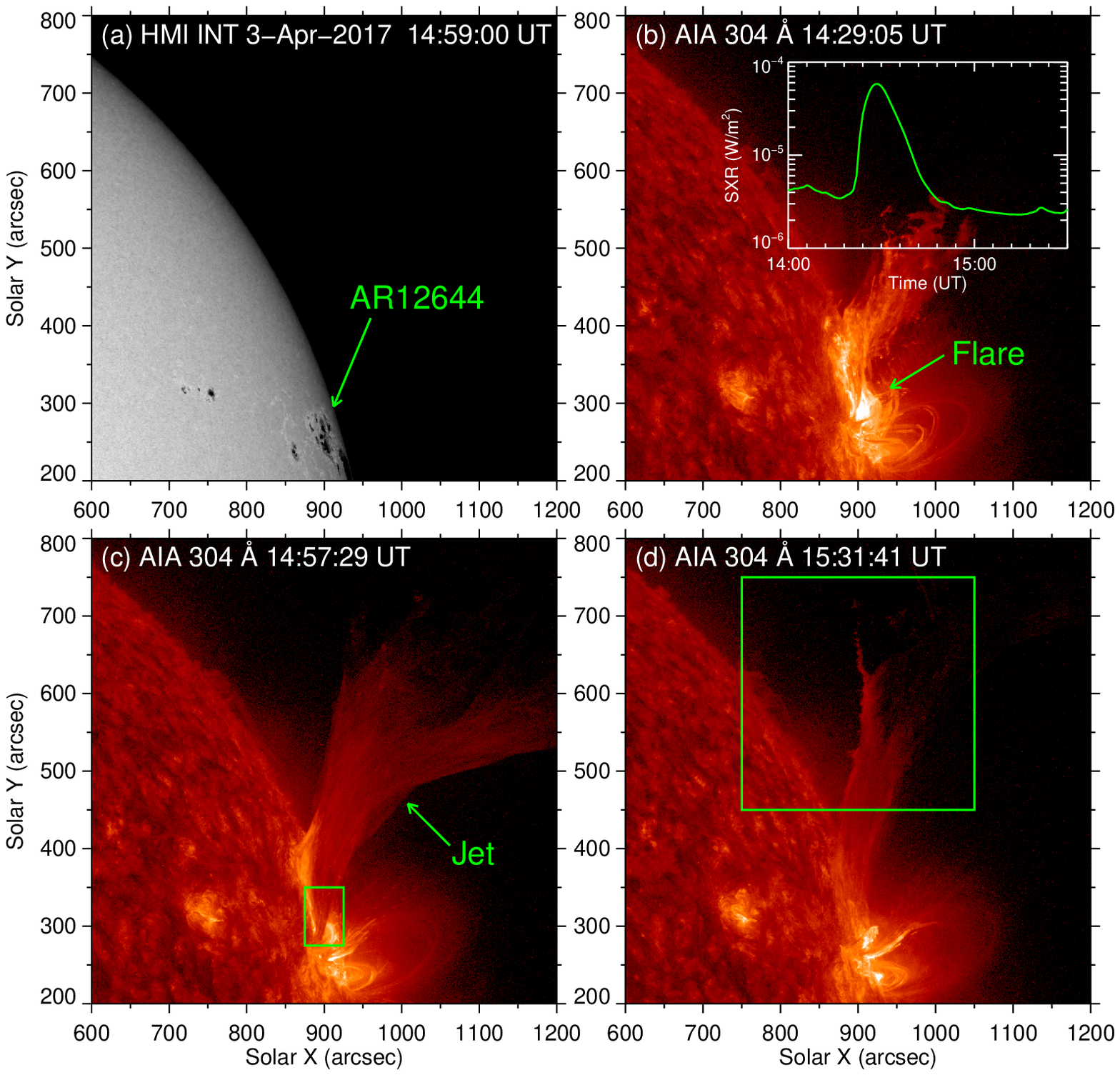}
\caption{HMI continuum intensity (panel (a)) and AIA 304 {\AA} images
(panels (b)-(d)) displaying the overview of the jet on 2017 April 3.
In panel (b), the green curve displays the variation of the
\emph{GOES} soft X-ray 1$-$8 {\AA} flux. The green rectangle in panel (c)
outlines the field-of-view (FOV) of Figure 2 and the green square
in panel (d) outlines the FOV of Figure 3.
\label{fig1}}
\end{figure*}

\begin{figure*}
\centering
\includegraphics
[bb=30 200 530 577,clip,angle=0,width=1.0\textwidth]{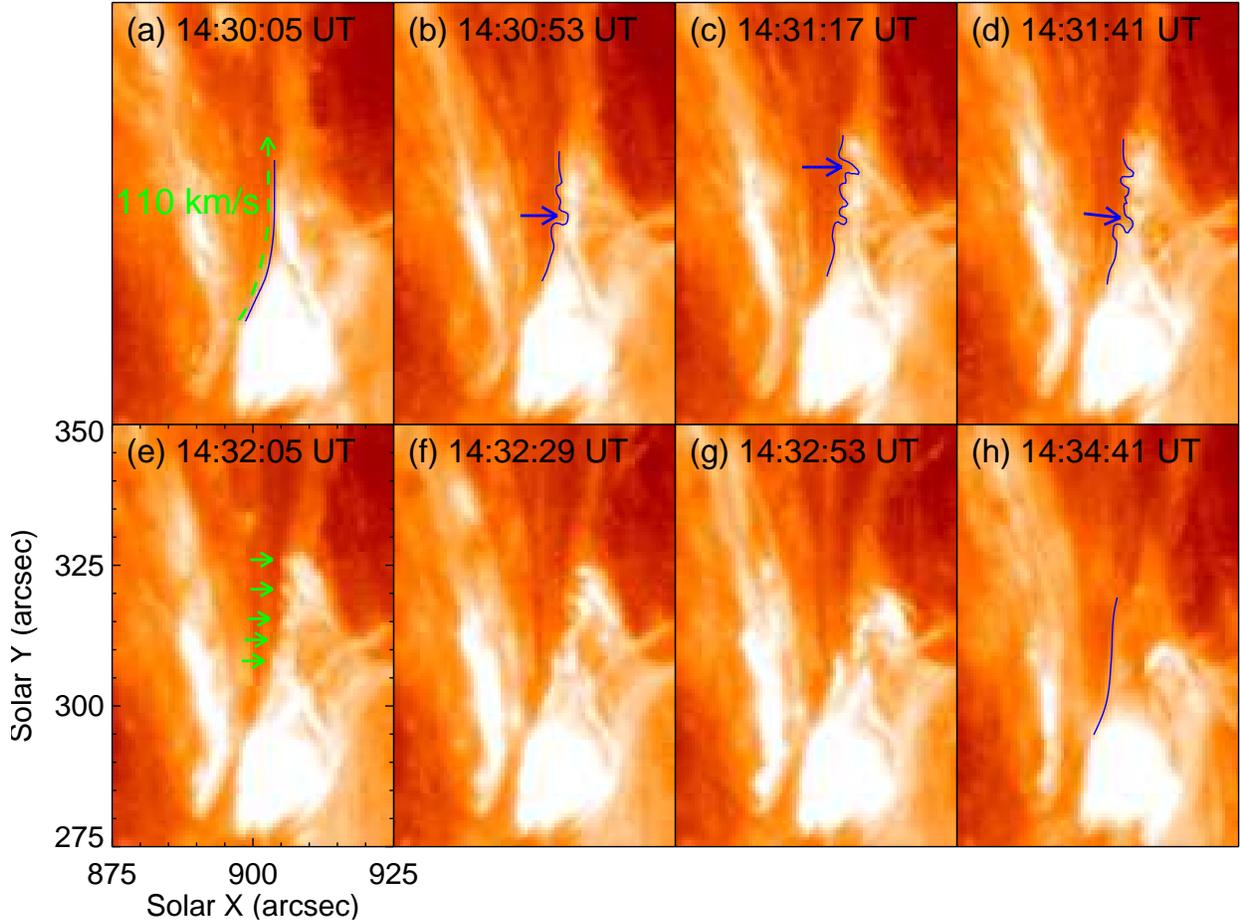}
\caption{AIA 304 {\AA} images showing the KHI in the
upstream regime of the jet. The green dashed arrow in panel (a)
displays the direction of the flow. In panels (a)-(d) and (h),
the blue curves denote the left boundary of the jet. The green
arrows in panel (e) point out five vortex-like structures caused
by the KHI.
\label{fig2}}
\end{figure*}

\begin{figure*}
\centering
\includegraphics
[bb=25 115 570 690,clip,angle=0,width=1.0\textwidth]{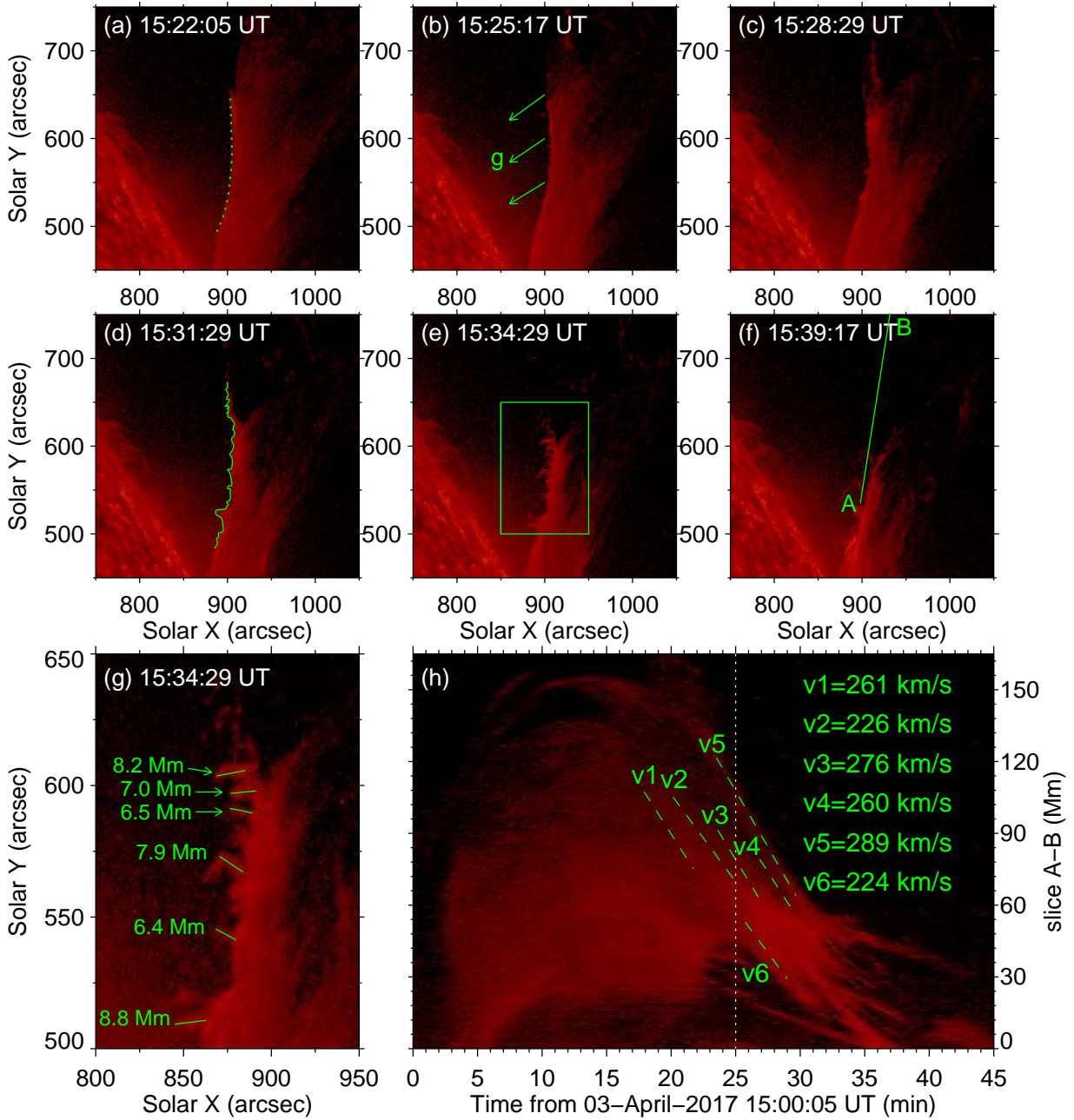}
\caption{AIA 304 {\AA} images (panels (a)-(f)) displaying the
development of the vortex-like structures triggered by the falling material.
The green curves in panels (a) and (d) denote the left boundary which
changes from being smooth into a sawtooth pattern. The green
arrows in panel (b) display the directions of gravity at the points
where arrows start. Panel (g) shows the
expanded view of the area outlined by the green rectangle in panel (e),
and the distortion values (6.4 $-$ 8.8 Mm) of six vortex-like
structures are marked. Panel (h) displays the space-time plot
along the slice ``A-B" as marked in panel (f). The velocities of
selected representative bright structures are displayed. The white
dotted line indicates the occurrence time of the vortex-like structures.
An animation (Movie1) of this figure is available.
\label{fig3}}
\end{figure*}

\begin{figure*}
\centering
\includegraphics
[bb=80 260 533 620,clip,angle=0,width=1.0\textwidth]{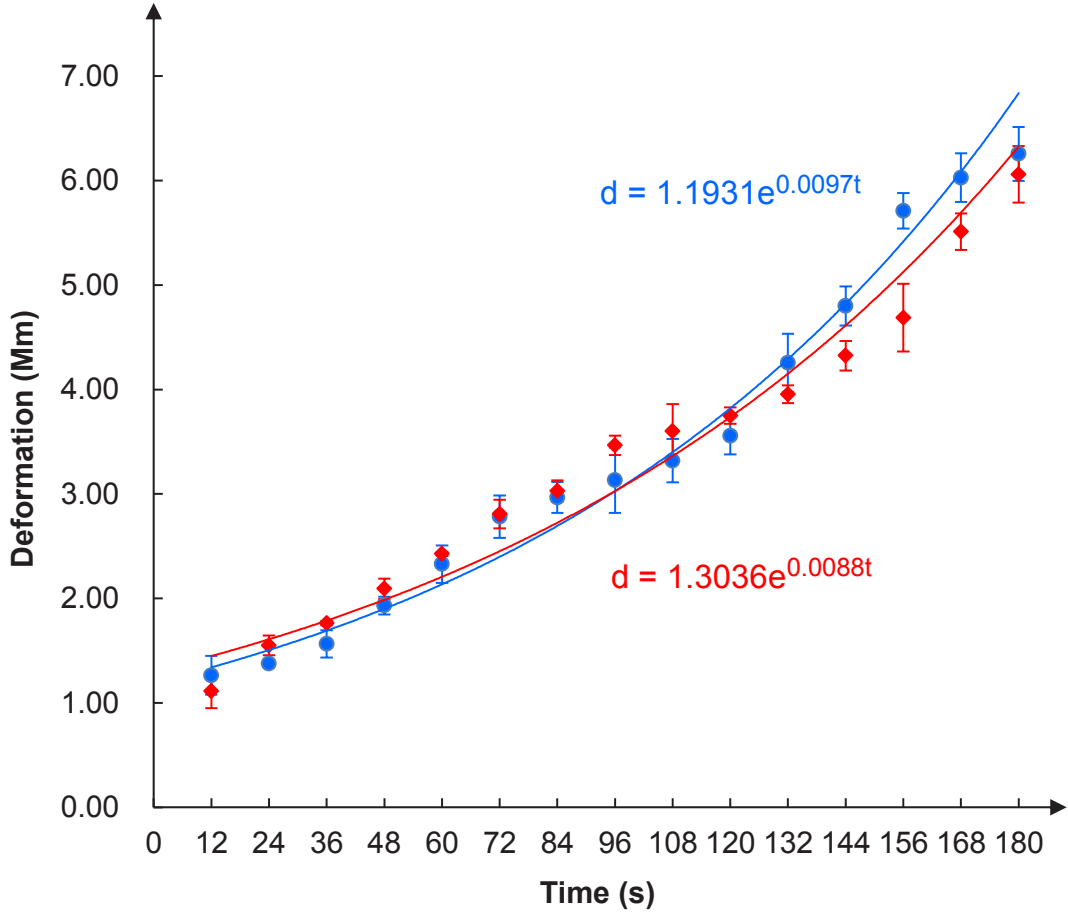}
\caption{Growth rates of the vortex-like structures.
We chose two structures in Figure 3(g) and measured their deformations
over time for five times. The blue and red points plot the
average values and the error bars indicate the standard deviations.
The blue and red curves denote corresponding fitted curves, and
the fitted equations are displayed beside.
\label{fig4}}
\end{figure*}

\begin{figure*}
\centering
\includegraphics
[bb=20 414 565 780,clip,angle=0,width=1.0\textwidth]{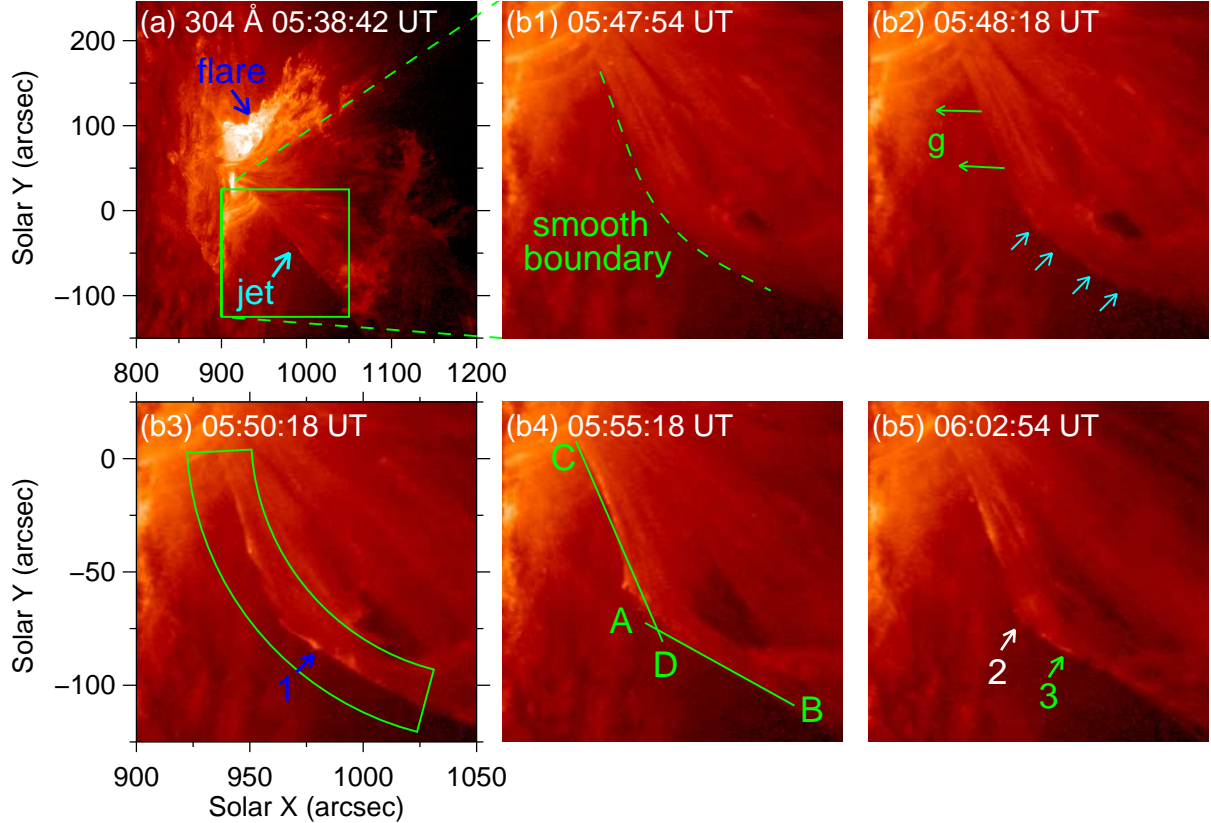}
\caption{AIA 304 {\AA} images displaying the vortex-like structures
in the downstream regime of the jet on 2016 July 23. Panel (a):
Appearances of the M5.5 class flare and the jet. Panels (b1)$-$(b5):
Development of the vortex-like structures at the south boundary
of the jet. The green arrows in panel (b2) display the directions
of gravity at the points where arrows start. The cyan arrows in
panel (b2) indicate several ripple structures. The green enclosed area in
panel (b3) outlines the FOV of Figure 7. The numbered arrows in
(b3) and (b5) indicate the structures which are displayed in Figure 7
with the same colors. In panel (b4), lines ``A$-$B" and ``C$-$D"
display the cross-cut positions used to obtain the stack plots shown
in Figure 6. An animation (Movie2) of this figure is available.
\label{fig5}}
\end{figure*}

\begin{figure*}
\centering
\includegraphics
[bb=30 45 565 421,clip,angle=0,width=1.0\textwidth]{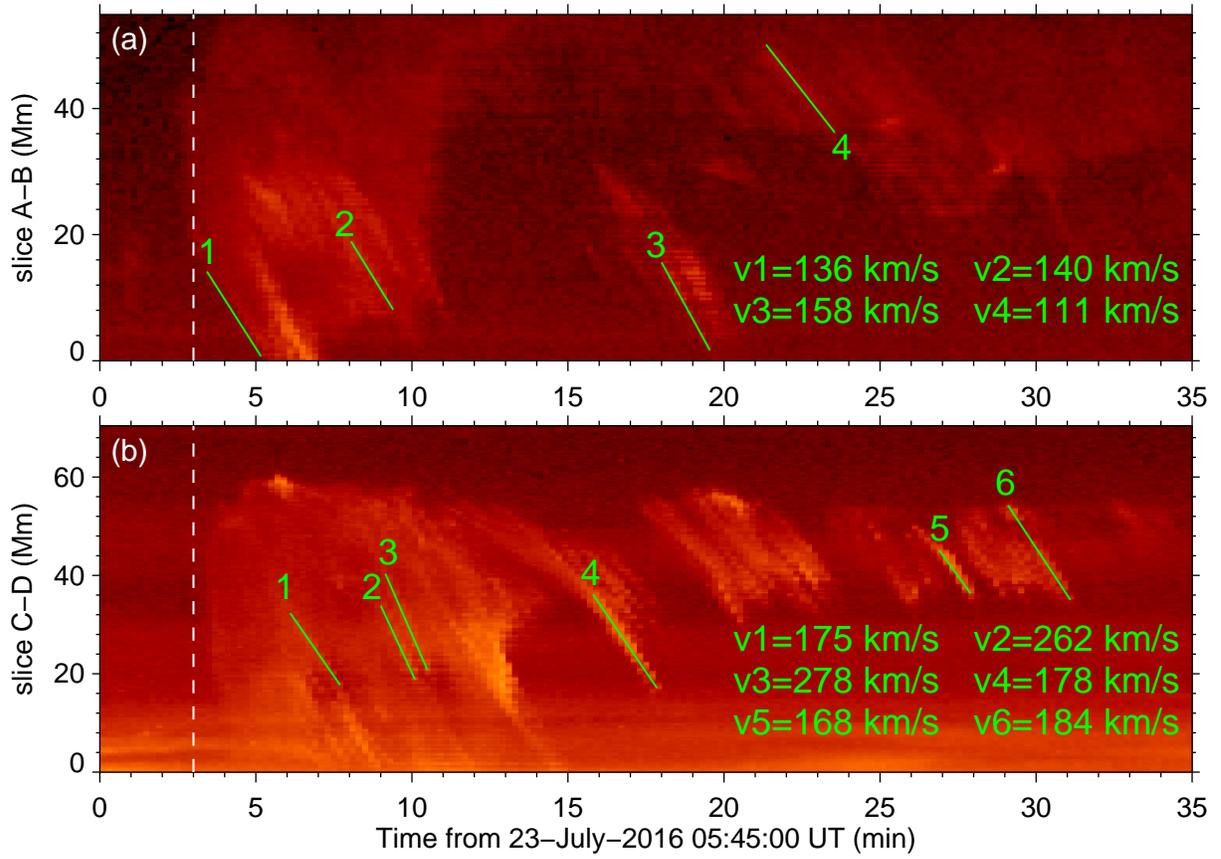}
\caption{Temporal evolution at the positions of lines
``A$-$B" and ``C$-$D" in Figure 5(b4). The velocities of
selected representative structures are displayed.
\label{fig6}}
\end{figure*}

\begin{figure*}
\centering
\includegraphics
[bb=103 146 491 697,clip,angle=0,width=0.9\textwidth]{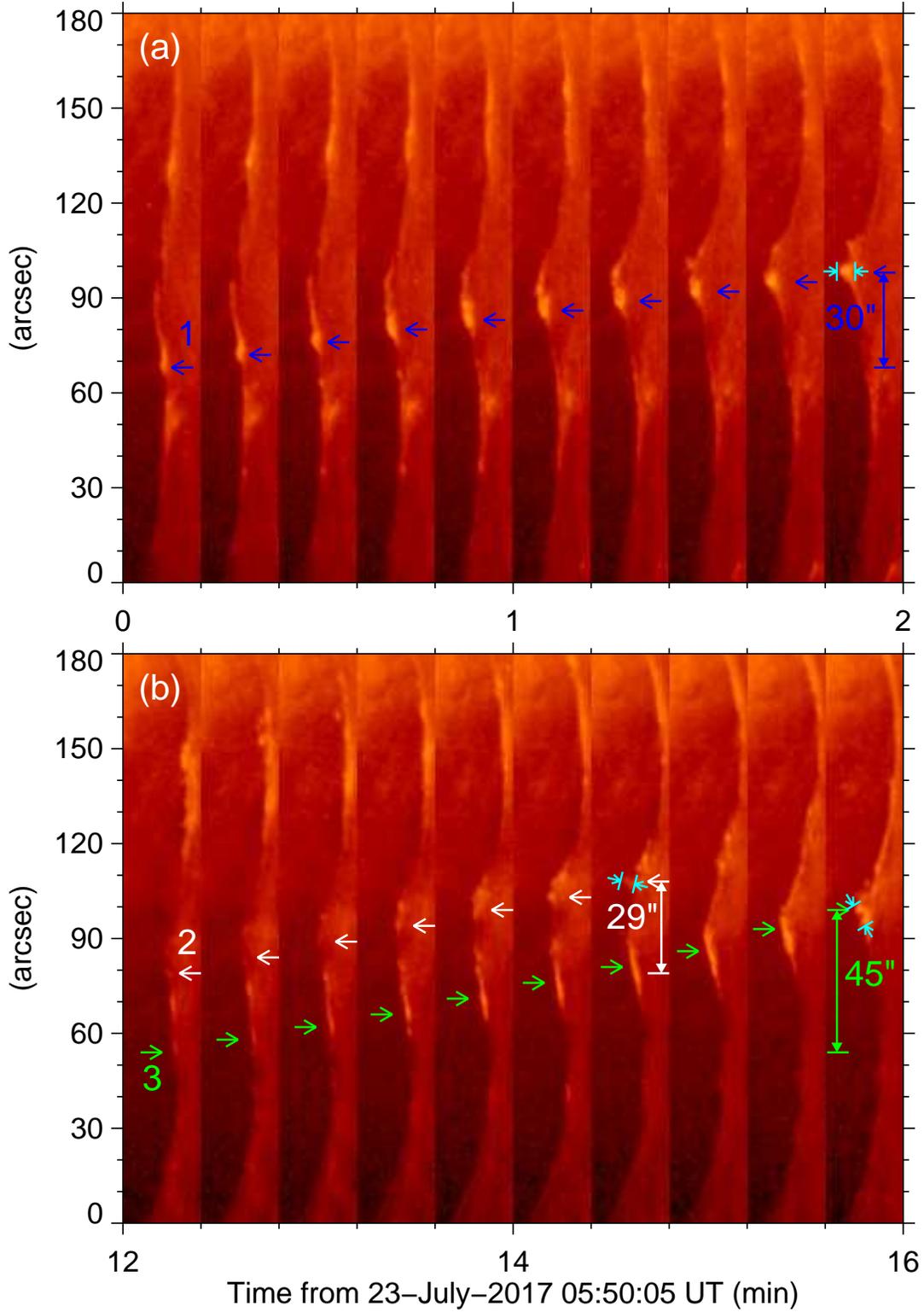}
\caption{Development of south boundary of the jet. The arrows
with same color denote the change of the location of the same
vortex-like structure caused by the falling material. The diameter of each
structure in the end is indicated by the cyan lines and arrows.
\label{fig7}}
\end{figure*}

\end{document}